\documentclass{article}
\usepackage{ijcai18}

\usepackage{times}
\usepackage{xcolor}
\usepackage{soul}
\usepackage[utf8]{inputenc}
\usepackage[small]{caption}
\usepackage{amsfonts}
\usepackage{amsmath}
\usepackage{amssymb}
\usepackage{subfigure}
\usepackage{graphicx}
\usepackage{multirow}
\usepackage{adjustbox}
\usepackage{array}
\usepackage{supertabular,booktabs}

\usepackage{enumitem}
\setenumerate[1]{itemsep=0pt,partopsep=0pt,parsep=\parskip,topsep=5pt}
\setitemize[1]{itemsep=0pt,partopsep=0pt,parsep=\parskip,topsep=5pt}
\setdescription{itemsep=0pt,partopsep=0pt,parsep=\parskip,topsep=5pt}
\usepackage{setspace}

\title{Where to Go Next: A Spatio-temporal LSTM model for \\Next POI Recommendation}

\author{
Pengpeng Zhao$^1$,
Haifeng Zhu$^1$,
Yanchi Liu$^2$,
Zhixu Li$^1$,
Jiajie Xu$^1$,
Victor S. Sheng$^3$,
\\
$^1$ School of Computer Science and Technology, Soochow University, Suzhou, 215006, China \\
$^2$ Management Science and Information Systems, Rutgers University, Piscataway, NJ, USA\\
$^3$ Department of Computer Science, University of Central Arkansas, Conway, USA  \\
ppzhao@suda.edu.cn,
second@email.address,
hfzhu@stu.suda.edu.cn,
\{zhixuli, xujj\}@suda.edu.cn,
ssheng@uca.edu
}

\begin{document}

\maketitle

\begin{abstract}
Next Point-of-Interest (POI) recommendation is of great value for both location-based service providers and users. Recently Recurrent Neural Networks (RNNs) have been proved to be effective on sequential recommendation tasks.
However, existing RNN solutions rarely consider the spatio-temporal intervals between neighbor check-ins, which are essential for modeling user check-in behaviors in next POI recommendation.
In this paper, we propose a new variant of LSTM, named ST-LSTM, which implements time gates and distance gates into LSTM to capture the spatio-temporal relation between successive check-ins.
Specifically,
one time gate and one distance gate are designed to control short-term interest update, and another time gate and distance gate are designed to control long-term interest update.
Furthermore, to reduce the number of parameters and improve efficiency, we further integrate coupled input and forget gates with our proposed model.
Finally, we evaluate the proposed model using four real-world datasets from various location-based social networks. Our experimental results show that our model significantly outperforms the state-of-the-art approaches for next POI recommendation.
\end{abstract}

\section{Introduction}
Recent years have witnessed the rapid growth of location-based social network services, such as Foursquare, Facebook Places, Yelp and so on. These services have attracted many users to share their locations and experiences 
with massive amounts of geo-tagged data accumulated, e.g., 55 million users generated more than 10 billion check-ins on Foursquare until December 2017.
These online footprints (or check-ins) provide an excellent opportunity to understand users' mobile behaviors.
For example, we can analyze and predict where a user will go next based on historical footprints. 
Moreover, such analysis can benefit POI holders to predict the customer arrival in the next time period.

\begin{figure} 
\centering
\setlength{\belowcaptionskip}{-0.4cm}
\begin{minipage}{0.4\linewidth}
    \subfigure[Language Modeling]{
      \label{experiment1:a} 
      \includegraphics[width=\textwidth]{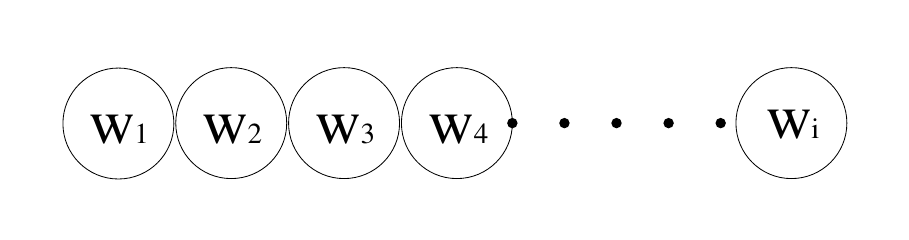}}
    \subfigure[Next Basket RS]{
      \label{experiment1:b} 
      \includegraphics[width=3.8cm]{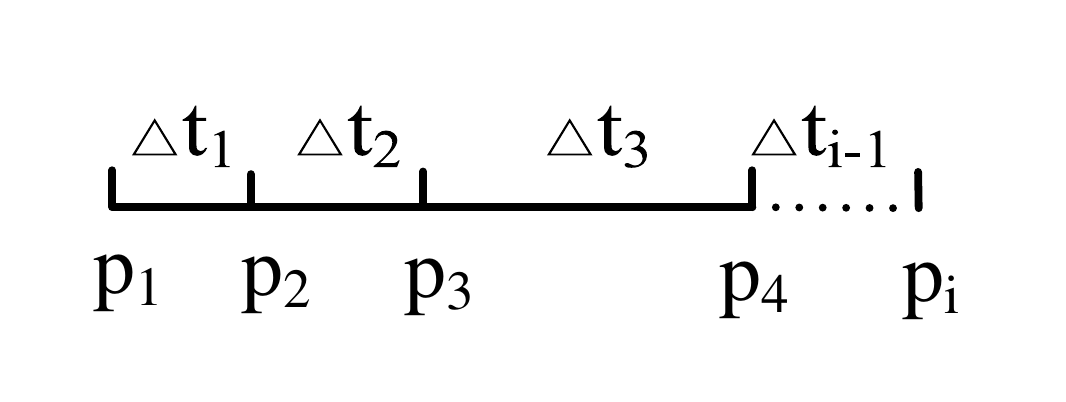}}
\end{minipage}
\begin{minipage}{0.55\linewidth}
	\subfigure[Next POI RS]{
      \label{experiment2:a} 
      \includegraphics[width=\textwidth, height=3cm]{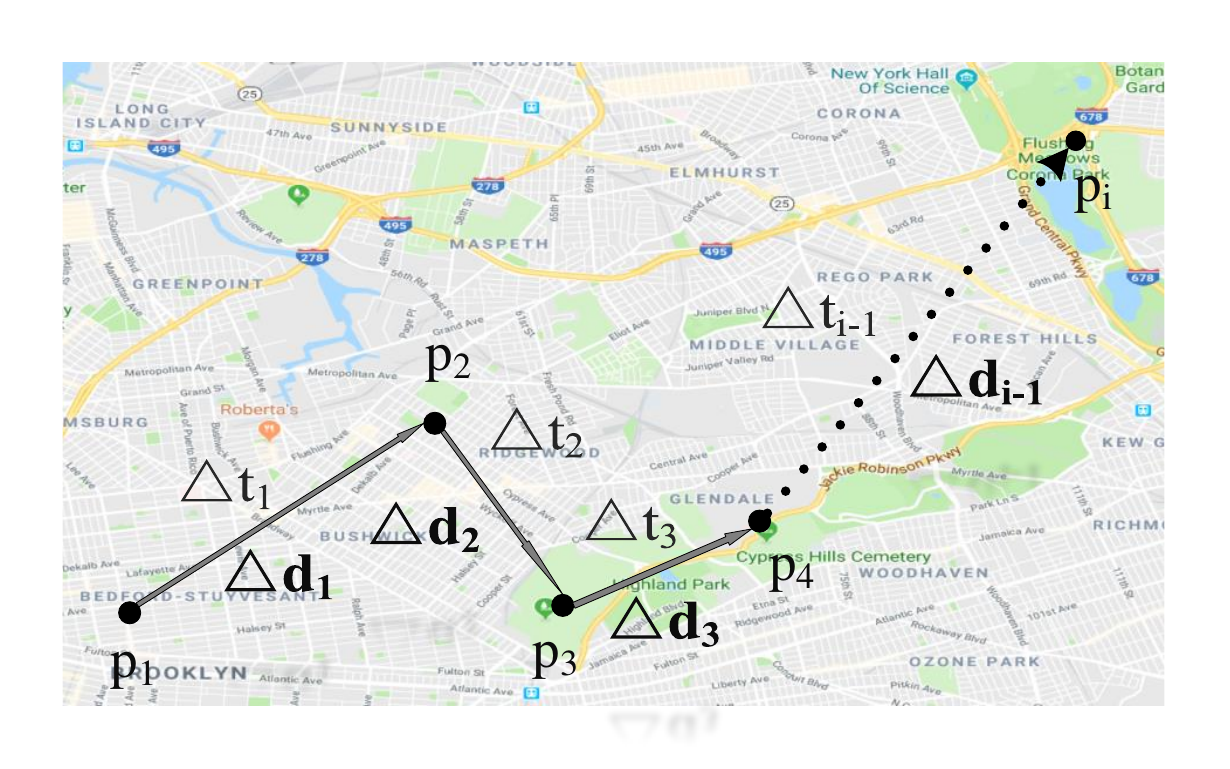}}

   \label{experiment2}
\end{minipage}
\vspace{-0.2cm}
\caption{$w_i$ in (a) represents the $i$-th word. In (b), $p_i$ represents the $i$-th item and $\triangle t$ is time interval between two neighbor items. In (c), $\triangle d$ further represents distance interval between two successive check-ins.}
\label{f:intro}
\end{figure}

In the literature, approaches like latent factor model and Markov chain have been widely applied for sequential data analysis and recommendation.
\cite{rendle2010factorizing} proposed Factorizing Personalized Markov Chain (FPMC), which bridges matrix factorization and Markov chains together, for next-basket recommendation.
\cite{cheng2013you} extended FPMC to embed personalized Markov chain and user movement constraint for next POI recommendation. \cite{he2016inferring} proposed a unified tensor-based latent model to capture the successive check-in behavior by exploring the latent pattern-level preference for each user.
Recently, Recurrent Neural Networks (RNNs) have been successfully employed on modeling sequential data and become state-of-the-art methods. 
\cite{hidasi2015session} focused on RNN solutions for session-based recommendation task, where no user id exists, and recommendations are made only on short session data.
\cite{zhu2017next} proposed a variant of Long-Short Term Memory network (LSTM), called Time-LSTM, to equip LSTM with time gates to model time intervals for next item recommendation.

However, none of the above recommendation methods considers both time intervals and geographical distances between neighbor items, which makes next POI recommendation different from other sequential tasks such as language modeling and next-basket recommender system (RS).
As shown in Figure \ref{f:intro}, there is no spatio-temporal interval between neighbor words in language modeling, and there is no distance interval between neighbor items in next-basket RS, while there are time and distance intervals between neighbor check-ins in next POI recommendation.
Traditional RNN and its variants, e.g., LSTM and GRU, do well in modeling the order information of sequential data with constant intervals, but cannot model dynamic time and distance intervals as shown in Figure 1(c).
A recent work ST-RNN \cite{liu2016predicting} tried to extend RNN to model the temporal and spatial context for next location prediction.
In order to model temporal context, ST-RNN models multi-check-ins in a time window in each RNN cell.
Meanwhile, ST-RNN employs time-specific and distance-specific transition matrices to characterize dynamic time intervals and geographical distances, respectively.
Thus, ST-RNN can obtain improvement in the spatio-temporal sequential recommendation. However, there exist some challenges preventing ST-RNN from becoming the best solution for next POI recommendation.

First of all, ST-RNN may fail to model spatial and temporal relations of neighbor check-ins properly. ST-RNN adopts time-specific and distance-specific transition matrices between cell hidden states within RNN. Due to data sparsity,  ST-RNN cannot learn every possible continuous time intervals and geographical distances but partition them into discrete bins.
Secondly, ST-RNN is designed for short-term interests and not well designed for long-term interests.
\cite{jannach2015adaptation} reported that users' short-term and long-term interests are both significant on achieving the best performance. The short-term interest here means that recommended POIs should depend on recently visited POIs, and the long-term interest means that recommended POIs should depend on all historical visited POIs. Thirdly, it is hard to select the proper width of the time window for different applications in ST-RNN since it models not one element in each layer but multi-elements in a fixed time period.

To this end, in this paper, we propose a new recurrent neural network model, named ST-LSTM, to model users' sequential visiting behaviors. Time intervals and distance intervals of neighbor check-ins are modeled by time gate and distance gate, respectively. Note that there are two time gates and two distance gates in the ST-LSTM model.
One pair of time gate and distance gate is designed to exploit time and distance intervals to capture the short-term interest, and the other is to memorize time and distance intervals to model the long-term interest.
Furthermore, enlightened by \cite{greff2017lstm}, we use the coupled input and forget gates to reduce the number of parameters, making our model more efficient.
Experimental results on four real-world datasets show ST-LSTM significantly improves next POI recommendation performance.

To summarize, our contributions are listed as follows.
\begin{itemize}
\item To the best of our knowledge, this is the first work that models spatio-temporal intervals between check-ins under LSTM architecture to learn user's visiting behavior for the next POI recommendation.
\item A ST-LSTM model is proposed to incorporate carefully designed time gates and distance gates to capture the spatio-temporal interval information between check-ins. As a result, ST-LSTM well models user's short-term and long-term interests simultaneously.
\item  Experiments on four large-scale real-world datasets are conducted to evaluate the performance of our proposed model. Our experimental results show that our method outperforms state-of-the-art methods.
\end{itemize}

\section{Related Work}

In this section, we discuss related work from two aspects, which are POI recommendation and leveraging neural networks for recommendation.

\subsection{POI Recommendation}
Different from traditional recommendations (e.g., movie recommendation, music recommendation), POI recommendation is characterized by geographic information and no explicit rating information \cite{ye2011exploiting,lian2014geomf}.
Moreover, additional information, such as social influence, temporal information, review information, and transition between POIs, has been leveraged for POI recommendation.  
\cite{ye2011exploiting} integrated the social influence with a user-based Collaborative Filtering (CF) model and modeled the geographical influence by a Bayesian model. 
\cite{yuan2013time} utilized the temporal preference to enhance the efficiency and effectiveness of the solution.  \cite{kurashima2013geo} proposed a topic model, in which a POI is sampled based on its topics and the distance to historical visited POIs of a target user. 
\cite{liu2016unified} exploited users' interests and their evolving sequential preferences with temporal interval assessment to recommend POI in a specified time period. 

Next POI recommendation, as a natural extension of general POI recommendation, is recently proposed and has attracted great research interest.
Research has shown that the sequential influence between successive check-ins plays a crucial role in next POI recommendation since human movement exhibits sequential patterns.
A tensor-based model, named FPMC-LR, was proposed by integrating the first-order Markov chain of POI transitions and distance constraints for next POI recommendation \cite{cheng2013you}.
\cite{he2016inferring} further proposed a tensor-based latent model considering the influence of user's latent behavior patterns, which are determined by the contextual temporal and categorical information.
\cite{feng2015personalized} proposed a personalized ranking metric embedding method (PRME) to model personalized check-in sequences for next POI recommendation.
\cite{xie2016learning} proposed a graph-based embedding learning approach, named GE, which utilize bipartite graphs to model context factors in a unified optimization framework.

\subsection{Neural Networks for Recommendation}
Neural networks are not only naturally used for feature learning to model various features of users or items,  
but also explored as a core recommendation model to simulate nonlinear, complex interactions between users and items \cite{wang2014improving,zhang2016collaborative}. 
\cite{zheng2016neural} further improved it with an autoregressive method. 
\cite{yang2017bridging} proposed a deep neural architecture named PACE for POI recommendation, which utilizes the smoothness of semi-supervised learning to alleviate the sparsity of collaborative filtering.
\cite{yang2017neural} jointly modeled a social network structure and users' trajectory behaviors with a neural network model named JNTM.
\cite{zhang2017next} tried to learn user's next movement intention and incorporated different contextual factors to improve next POI recommendation.
\cite{zhu2017next} proposed a Time-LSTM model and two variants, which equip LSTM with time gates to model time intervals for next item recommendation.

A recent work proposed a model named ST-RNN, which considers spatial and temporal contexts to model user behavior for next location prediction, is closely related to our work \cite{liu2016predicting}.
However, our proposed ST-LSTM model differs significantly from ST-RNN in two aspects. 
First, ST-LSTM equips the LSTM model with time and distance gates while ST-RNN adds spatio-temporal transition matrices to the RNN model.
Second, ST-LSTM well models time and distance intervals between neighbor check-ins to extract long-term and short-term interests. However, ST-RNN recommends next POI depending only on POIs in the nearest time window which may be hard to distinguish short-term and long-term interests.

\section{Preliminaries}

In this section, we first give the formal problem definition of next POI recommendation, 
and then briefly introduce LSTM.

\subsection{Problem Formulation}
Let $\mathbb{U} = \left\{ {{u_1},{u_2}, \ldots, {u_M}} \right\}$ be the set of $M$ users and $\mathbb{V} = \left\{ {{v_1},{v_2}, \ldots, {v_N}} \right\}$ be the set of $N$ POIs.
For user $u$, she has a sequence of historical POI visits up to time $t_{i-1}$ represented as $H_i^u=\{v_{t_1}^u,v_{t_2}^u,\cdot\cdot\cdot,v_{t_{i-1}}^u\}$, where $v_{t_{i}}^u$ means user $u$ visit POI $v$ at time $t_i$.
The goal of next POI recommendation is to recommend a list of unvisited POIs for a user to visit next at time point $t_i$.
Specifically, a higher prediction score of a user $u$ to an unvisited POI ${v_j}$ indicates a higher probability that the user $u$ would like to visit the POI ${v_j}$ at time $t_i$. According to prediction scores, we can recommend top-$k$ POIs to user $u$.

\subsection{LSTM}
LSTM \cite{hochreiter1997long}, a variant of RNN, is capable of learning short and long-term dependencies.
LSTM has become an effective and scalable model for sequential prediction problems, and many improvements have been made to the original LSTM architecture. We use the basic LSTM model in our approach for the concise and general purpose, and it is easy to extend to other variants of LSTM.
The basic update equations of LSTM are as follows:
\begin{flalign}
\label{eq1}
\ \ \ \ \ \ \ \ \ \ i_{t} &= \sigma (W_{i}[h_{t-1},x_{t}]+b_{i}),&
\end{flalign}
\vspace{-0.4cm}
\begin{flalign}
\label{eq2}
\ \ \ \ \ \ \ \ \ \ f_{t}& = \sigma (W_{f}[h_{t-1},x_{t}]+b_{f}),&
\end{flalign}
\vspace{-0.4cm}
\begin{flalign}
\label{c_t-1}
\ \ \ \ \ \ \ \ \ \ \widetilde{c_{t}}&=\tanh (W_{c}[h_{t-1},x_{t}]+b_{c}),&
\end{flalign}
\vspace{-0.4cm}
\begin{flalign}
\label{c_t}
\ \ \ \ \ \ \ \ \ \ c_{t}&=f_{t} \odot c_{t-1}+i_{t} \odot \widetilde{c_{t}},&
\end{flalign}
\vspace{-0.4cm}
\begin{flalign}
\label{eq5}
\ \ \ \ \ \ \ \ \ \ o_{t}& = \sigma (W_{o}[h_{t-1},x_{t}]+b_{o}),&
\end{flalign}
\vspace{-0.4cm}
\begin{flalign}
\label{eq6}
\ \ \ \ \ \ \ \ \ \ h_{t}&=o_{t} \odot \tanh (c_{t}),&
\end{flalign}
where $i_{t}$, $f_{t}$, $o_{t}$ represent the input, forget and output gates of the $t$-th object, deciding what information  to store, forget and output, respectively. $c_{t}$ is the cell activation vector representing cell state, which is the key to LSTM. $x_{t}$ and $h_{t}$ represent the input feature vector and the hidden output vector, respectively. $\sigma$ represents a sigmoid layer to map the values between 0 to 1, where 1 represents ``complete keep this" while 0 represents ``completely get rid of this".
$W_{i}$, $W_{f}$, $W_{o}$ and $W_{c}$ are the weights of gates. $b_{i}$, $b_{f}$, $b_{o}$ and $b_{c}$ are corresponding biases. And $\odot$ represents for the element-wise (Hadamard) product.
The update of cell state $c_{t}$ has two parts. The former part is the previous cell state $c_{t-1}$ that is controlled by forget gate $f_{t}$, and the latter part is the new candidate value scaled by how much we decided to add state value.

\section{Our Approach}

In this section, we first propose a spatio-temporal LSTM model, ST-LSTM, which utilizes time and distance intervals to model user's short-term interest and long-term interest simultaneously.
Then, we improve ST-LSTM with coupled input and output gates for efficiency.

\subsection{Spatio-temporal LSTM}

When using LSTM for next POI recommendation, $x_t$ represents user's last visited POI, which can be exploited to learn user's short-term interest. While $c_{t-1}$ contains the information of user's historical visited POIs, which reflect user's long-term interest.
However, how much the short-term interest determines where to go next heavily depends on the time interval and the geographical distance between the last POI and the next POI.
Intuitively, a POI visited long time ago and long distance away has little influence on next POI, and vice versa. 
In our proposed ST-LSTM model, we use time gate and distance gate to control the influence of the last visited POI on next POI recommendation.
Furthermore, the time gate and the distance gate can also help to store time and distance intervals in cell state $c_{t}$, which memorizes user's long-term interest.
In this way, we utilize time and distance intervals to model user's short-term interest and long-term interest simultaneously.

\begin{figure}
\setlength{\abovecaptionskip}{0.2cm}
\setlength{\belowcaptionskip}{-0cm}
\footnotesize\centerline{\includegraphics[width=0.5\textwidth]{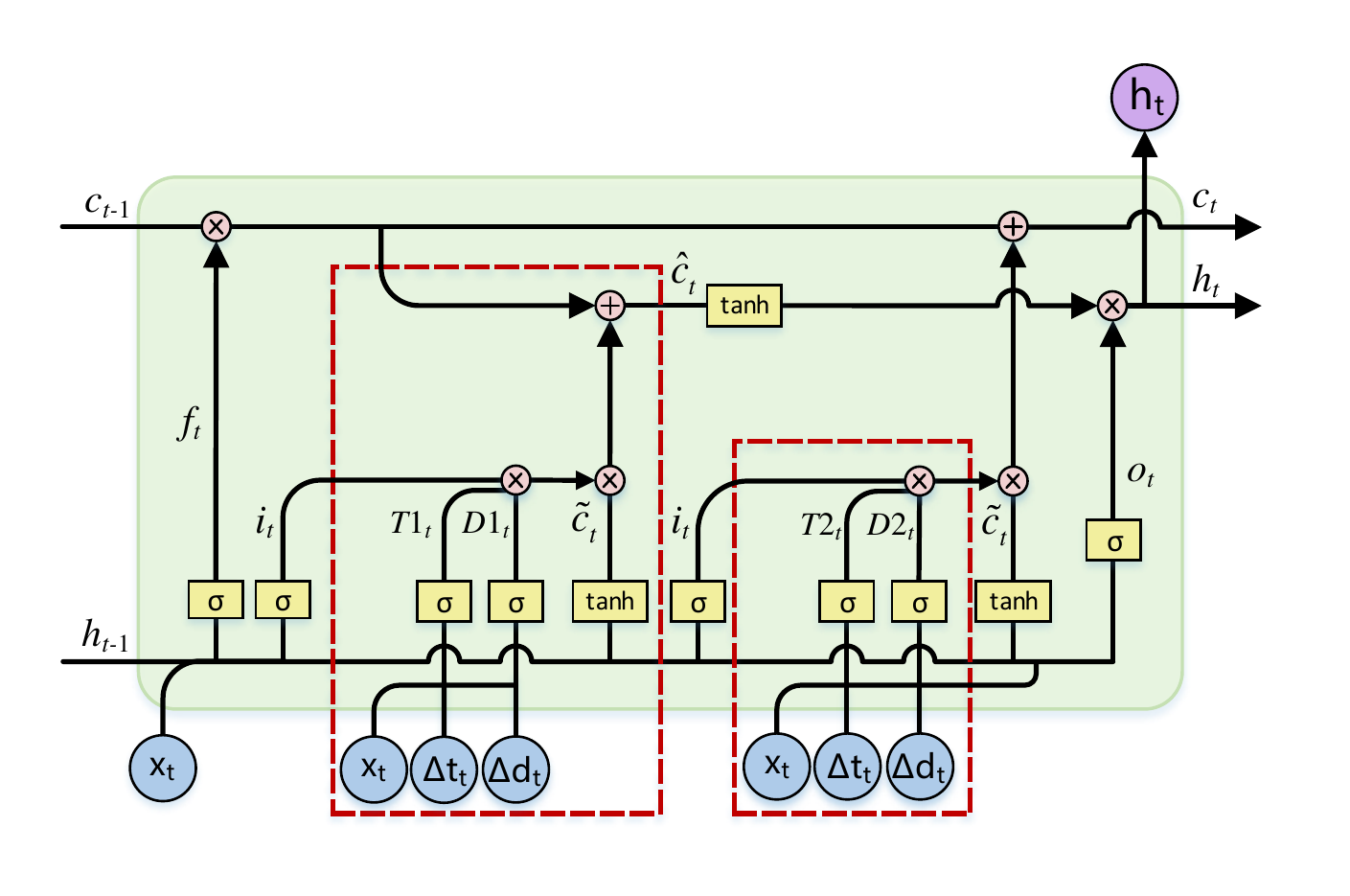} }
\vspace{0.2cm}
\caption{ST-LSTM  has two time gates and two distance gates, i.e., $T1_t$, $T2_t$, $D1_t$ and $D2_t$. $T1_t$ and $D1_t$ are designed to model time and distance intervals for short-term interests while $T2_t$ and $D2_t$ are to model time and distance intervals for long-term interest.}
\label{model:1}
\end{figure}

As shown in two dotted red rectangles in Figure \ref{model:1}, we add two time gates and two distance gates to LSTM, denoted as $T1_t$, $T2_t$, $D1_t$ and $D2_t$ respectively.
$T1_t$ and $D1_t$ are used to control the influence of the latest visited POI on next POI, and
$T2_t$ and $D2_t$ are used to capture time and distance intervals to model user's long-term interest. Based on LSTM, we add equations for time gates and distance gates as follows:
\begin{spacing}{1}
\begin{flalign}
\begin{split}
\label{eq7}
\ \ \ \ T1_t = & \sigma(x_t W_{xt_1}+ \sigma(\triangle t_t W_{t_1})+b_{t_1}), \\
\ \ \ \ \ & s.t. W_{xt_1} \leq 0
\end{split}&
\end{flalign}
\vspace{-0.2cm}
\begin{flalign}
\label{eq8}
\ \ \ \ T2_t& = \sigma(x_t W_{xt_2}+ \sigma(\triangle t_t W_{t_2})+b_{t_2}),&
\end{flalign}
\vspace{-0.4cm}
\begin{flalign}
\begin{split}
\label{eq9}
\ \ \ \ D1_t = &\sigma(x_t W_{xd_1}+ \sigma(\triangle d_t W_{d_1})+b_{d_1}), \\
\ \ \ \ \ &s.t. W_{xd_1} \leq 0
\end{split}&
\end{flalign}
\vspace{-0.2cm}
\begin{flalign}
\label{eq10}
\ \ \ \ D2_t& = \sigma(x_t W_{xd_2}+ \sigma(\triangle d_t W_{d_2})+b_{d_2}).&
\end{flalign}
\end{spacing}
We then modify Eq.~(4)-(6) to:
\begin{flalign}
\label{eq11}
\ \ \hat{c_t}& = f_t \odot c_{t-1} +i_t \odot T1_t \odot D1_t \odot \tilde{c_t},&
\end{flalign}
\vspace{-0.2cm}
\begin{flalign}
\label{eq12}
\ \ c_t& = f_t \odot c_{t-1} +i_t \odot T2_t \odot D2_t \odot \tilde{c_t},&
\end{flalign}
\vspace{-0.2cm}
\begin{flalign}
\label{eq13}
\ \ o_t& = \sigma (W_o[h_{t-1},x_t]+ \triangle t_t W_{to}+\triangle d_t W_{do}+b_o),&
\end{flalign}
\vspace{-0.2cm}
\begin{flalign}
\label{eq14}
\ \ h_t& = o_t \odot \tanh(\hat{c_t}),&
\end{flalign}
where
$\triangle t_t$ 
is the time interval and $\triangle  d_t$ is the distance interval.
Besides input gate  $i_t$, $T1_t$ can be regarded as an input information filter considering time interval, and
$D1_t$ can be regarded as another input information filter considering distance interval.
We add a new cell state $\hat{c_t}$ to store the result, then transfer to the hidden state $h_t$ and finally influences next recommendation. Along this line, $\hat{c_t}$ is filtered by time gate $T1_t$ and distance gate $D1_t$ as well as input gate $i_t$ on current recommendations.

Cell state $c_t$ is used to memory users general interest, i.e., long-term interest. We designed a time gate and a distance gate to control the cell state $c_t$ update. $T2_t$ first memorizes $\triangle t_t$ then transfers to $c_t$, further to $c_{t+1},c_{t+2},\cdots$. So $T2_t$ helps store $\triangle t_t$ to model user long-term interest. In the similar way, $D2_t$ memorizes $\triangle d_t$ and transfers to cell state $c_t$ to help model user long-term interest. In this way, $c_t$ captures user long-term interest by memorizing not only the order of user's historical visited POIs, but also the time and distance interval information between neighbor POIs. 
Modeling distance intervals can help capture user's general spatial interest, while modeling time intervals helps capture user's periodical visiting behavior.

Normally, a more recently visited POI with a shorter distance should have a larger influence on choosing next POI. To incorporate this knowledge in the designed gates, we add constraints $W_{xt1}\leq 0$ and $W_{xd1} \leq 0$ in Eq.~(\ref{eq7}) and Eq.~(\ref{eq9}).
Accordingly, if $\triangle t_t$ is smaller, $T1_t$ would be larger according to Eq.~(\ref{eq7}). In the similar way, if $\triangle d_t$ is shorter, $D1_t$ would be larger according to Eq.~(\ref{eq9}).
For example, if time and distance intervals are smaller between $x_t$ and next POI, then $x_t$ better indicates the short-term interest, thus its influence should be increased.
If $\triangle t_t$ or $\triangle d_t$ is larger, $x_t$ would have a smaller influence on the new cell state $\hat{c}$.
In this case, the short-term interest is uncertain, so we should depend more on the long-term interests.
It is why we set two time gates and two distance gates to distinguish the short-term and long-term interests update.

\subsection{Variation of coupled input and forget gates}

\begin{figure}
\setlength{\belowcaptionskip}{-0.4cm}
\footnotesize\centerline{\includegraphics[width=0.5\textwidth]{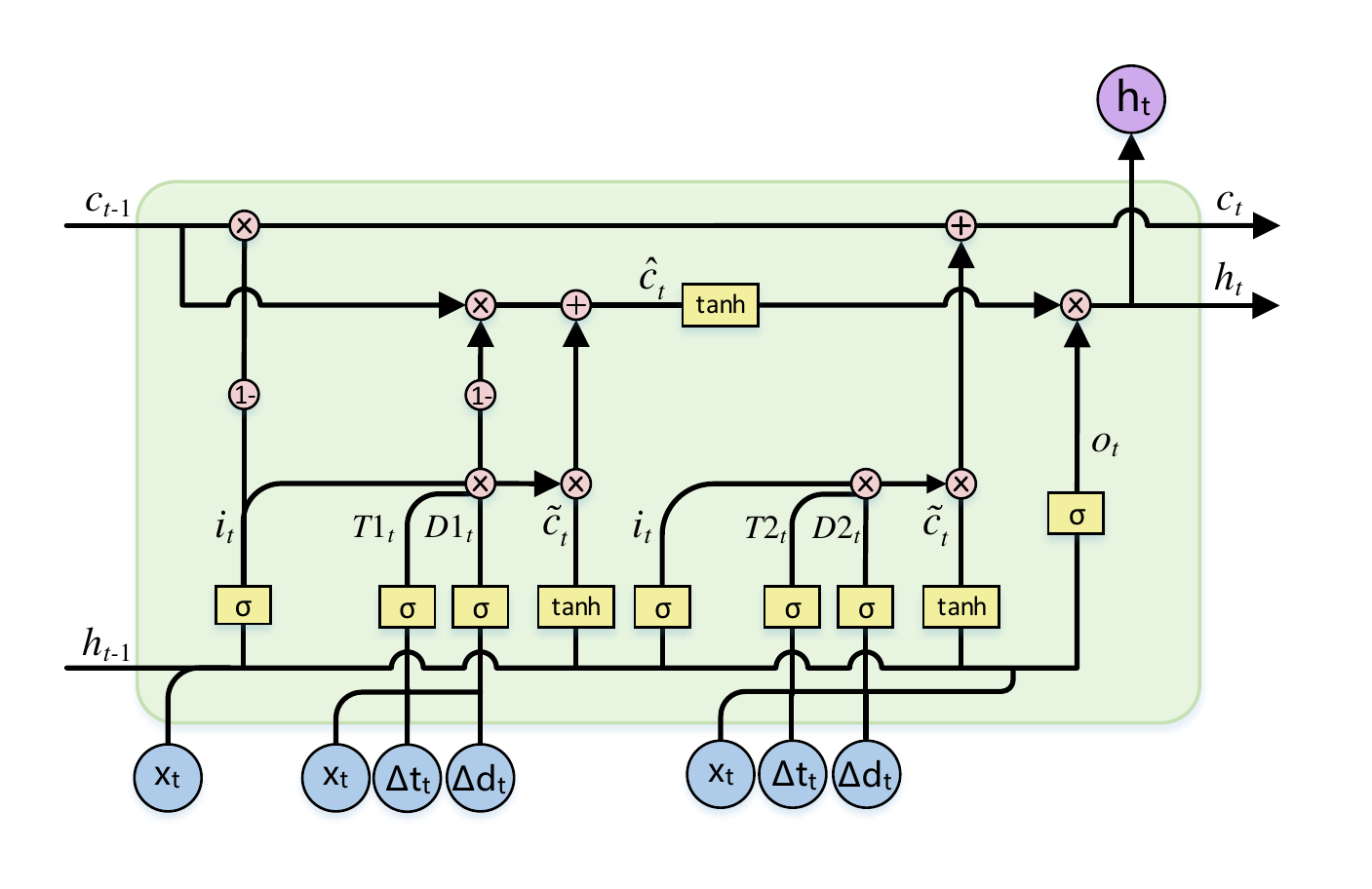} }
\vspace{-0cm}
\caption{A variant of ST-LSTM using coupled input and forget gates.}
\label{model:2}
\end{figure}

Enlightened by \cite{greff2017lstm}, we propose another version of ST-LSTM, named ST-CLSTM, to reduce the number of parameters and improve efficiency. ST-CLSTM uses coupled input and forget gates instead of separately deciding what to forget and what new information to add, as shown in Figure \ref{model:2}.
Specifically, we remove the forget gate, and modify Eq.~(\ref{eq11}) and Eq.~(\ref{eq12}) to:
\begin{flalign}
\begin{split}
\label{eq15}
\ \ \ \ \ \ \ \ \ \ \hat{c_t} = & (1-i_t \odot T1_t \odot D1_t) \odot c_{t-1} \\
\ \ \ \ \ & +i_t \odot T1_t \odot D1_t \odot \tilde{c_t},
\end{split}&
\end{flalign}
\vspace{-0.2cm}
\begin{flalign}
\label{eq16}
\ \ \ \ \ \ \ \ \ \ c_t& = (1-i_t) \odot c_{t-1} +i_t \odot T2_t \odot D2_t \odot \tilde{c_t}.&
\end{flalign}

Since time gate $T1_t$ and distance gate $D1_t$ are regarded as input filters, we replace the forget gate with $(1-i_t \odot T1_t \odot D1_t)$ in Eq.~(\ref{eq15}).
$T2_t$ and $D2_t$ are used to store time intervals and distance intervals respectively, thus we use $(1-i_t)$ in Eq.~(\ref{eq16}).

\subsection{Training}
The way we adapt our model to next POI recommendation is as follows.
Firstly we transform $H^u$ to $[(v_1^u,t_2^u-t_1^u,d(l_1,l_2)),(v_2^u,t_3^u-t_2^u,d(l_2,l_3)),\cdots,(v_n^u,t_q^u-t_n^u,d(l_n,l_q))]$. Then $x_t$ in ST-LSTM is equivalent to $v_t^u$, $\triangle t_t$ is equivalent to $t_{t+1}^u-t_t^u$, and $\triangle d_t$ is equivalent to $d(l_{t+1},l_t)$, where $d(\cdotp,\cdotp)$ is the function computing the distance between two geographical points.
Moreover, we make use of all users' behavioral histories for learning and recommendation. We leverage the mini-batch learning method, and train the model on users’ existing histories until convergence. The model output is a probability distribution on all POIs calculated by $h_t$ and $v_t^u$. And then we take a gradient step to optimize the loss based on the output and one-hot representations of $v_{t+1}^u$.

We use Adam, a variant of Stochastic Gradient Descent(SGD),  to optimize the parameters in ST-LSTM, which adapts the learning rate for each parameter by performing smaller updates for frequent parameters and larger updates for infrequent parameters. We use the projection operator described in \cite{rakhlin2012making} to meet the constraints $W_{t_1} \leq 0$ in Eq.~(\ref{eq7}) and $W_{d_1} \leq 0$ in Eq.~(\ref{eq9}). If we have $W_{t_1} > 0$ during the training process, we set $W_{t_1} = 0$. And parameter $W_{d_1}$ is set in the same way.

The computational complexity of learning LSTM models per weight and time step with the stochastic gradient descent (SGD) optimization technique is $O(1)$. Hence, the LSTM algorithm is very efficient, with an excellent update complexity of $O(W)$, where $W$ is the number of weights and can be calculated as
$W=n_c*n_c*4+n_i*n_c*4+n_c*n_o+n_c*3$, 
where $n_c$ is the number of memory cells, $n_i$ is the number of input units, and $n_o$ is the number of output units.
Similarly, ST-LSTM computational complexity is also $O(W)$ and can be calculated as
$W=n_c*n_c*5+n_i*n_c*8+n_c*n_o+n_c*9$
. The training time of our proposed model for $100$ rounds of training on four datasets after data cleaning is about $10$ minutes on GPU M6000.

\section{Experiments}
In this section, we conduct experiments to evaluate the performance of our proposed model ST-LSTM on four real-world datasets. We first briefly depict the datasets, followed by baseline methods. Finally, we present our experimental results and discussions.

\subsection{Dataset}
We use four public LBSNs datasets that have user-POI interactions of users and locations of POIs.
The statistics of the four datasets are listed in Table \ref{table:dataset}.
CA is a Foursquare dataset from users whose homes are in California, collected from January 2010 to February 2011 and used in \cite{gao2012gscorr}.
SIN is a Singapore dataset crawled from Foursquare used by \cite {yuan2013time}. Gowalla\footnote{http://snap.stanford.edu/data/loc-gowalla.html} and Brightkite\footnote{http://snap.stanford.edu/data/loc-brightkite.html} are two widely used LBSN datasets, which have been used in many related research papers. We eliminate users with fewer than 10 check-ins and POIs visited by fewer than 10 users in the four datasets. Then, we sorted each user's check-in records according to timestamp order, taking the first 70\% as training set, the remaining 30\% for the test set.

\begin{table}
\setlength{\abovecaptionskip}{0.1cm}
\centering
\caption{Statistics of the four datasets}
\vspace{0.1cm}
\small
\label{table:dataset}
\setlength{\belowcaptionskip}{-0.5cm}
\begin{tabular}{|l|llll|}
\hline
Dataset    & \#user & \#POI & \#Check-in & Density \\ \hline
CA &   49,005    &   206,097        &       425,691        &     0.004\%    \\ \hline
SIN       &   30,887     &    18,995   &        860,888     &       0.014\%   \\ \hline
Gowalla    &   18,737     &   32,510    &     1,278,274        &     0.209\%   \\ \hline
Brightkite &   51,406      &   772,967      &      4,747,288       &     0.012\%     \\ \hline
\end{tabular}
\vspace{-0.2cm}
\end{table}

\subsection{Baseline Methods}
We compare our proposed model ST-LSTM with seven representative methods for next POI recommendation.
\begin{itemize}
  \item \textbf{FPMC-LR} \cite{cheng2013you}: It combines the personalized Markov chains with the user movement constraints around a localized region. It factorizes the transition tensor matrices of all users and predicts next location by computing the transition probability.
  \item \textbf{PRME-G} \cite{feng2015personalized}: It utilizes the Metric Embedding method to avoid drawbacks of the MF. Specifically, it embeds users and POIs into the same latent space to capture the user transition patterns.
  \item \textbf{GE} \cite{xie2016learning}: It embeds four relational graphs (POI-POI, POI-Region, POI-Time, POI-Word) into a shared low dimensional space. The recommendation score is then calculated by a linear combination of inner products for these contextual factors.
  \item \textbf{RNN} \cite{zhang2014sequential}:
 This method leverages the temporal dependency in user's behavior sequence through a standard recurrent structure.
  \item \textbf{LSTM} \cite{hochreiter1997long} This is a variant of RNN model, which contains a memory cell and three multiplicative gates to allow long-term dependency learning.
  \item \textbf{GRU} \cite{cho2014learning}: This is a variant of RNN model, which is equipped with two gates to control the information flow.
  \item \textbf{ST-RNN} \cite{liu2016predicting}: Based on the standard RNN model, ST-RNN replaces the single transition matrix in RNN with time-specific transition matrices and distance-specific transition matrices to model spatial and temporal contexts.

\end{itemize}

\subsection{Evaluation Metrics}
To evaluate the performance of our proposed model ST-LSTM and compare with the seven baselines described above, we use two standard metrics Acc@K and Mean Average Precision (MAP). These two metrics are popularly used for evaluating recommendation results, such as \cite{liu2016predicting,he2016inferring,xie2016learning}. Note that for an instance in testing set,
Acc@K is 1 if the visited POI appears in the set of top-K recommendation POIs, and 0 otherwise. 
The overall Acc@K is calculated as the average value of all testing instances. 
In this paper, we choose K = \{1, 5, 10, 15, 20\} to illustrate different results of Acc@K. 

\subsection{Results and Discussions}

\begin{table*}
\setlength{\abovecaptionskip}{0.1cm}
\setlength{\belowcaptionskip}{-0.8cm}
\tiny
\centering
\caption{Evaluation of next POI recommendation in terms of Acc@K and MAP on four datasets}
\vspace{0.1cm}

\begin{supertabular}{p{1.25cm}<{\centering}|p{1.0cm}<{\centering}|p{1.0cm}<{\centering}|p{1.0cm}<{\centering}|p{1.0cm} <{\centering}|p{1.0cm}<{\centering}|p{1.0cm}<{\centering}|p{1.0cm} <{\centering}|p{1.0cm}<{\centering}}\hline

\multirow{2}{*}{} & \multicolumn{4}{c|}{\textbf{CA}} & \multicolumn{4}{c}{\textbf{SIN}} \\ \cline{2-9}
& \multicolumn{1}{c|}{\textbf{Acc@1}} & \multicolumn{1}{c|}{\textbf{Acc@5}} & \multicolumn{1}{c|}{\textbf{Acc@10}} & \multicolumn{1}{c|}{\textbf{MAP}} & \multicolumn{1}{c|}{\textbf{Acc@1}} & \multicolumn{1}{c|}{\textbf{Acc@5}} & \multicolumn{1}{c|}{\textbf{Acc@10}} & \multicolumn{1}{c}{\textbf{MAP}} \\ \hline
\multicolumn{1}{l|}{FPMC-LR} &0.0378 &0.0493 &0.0784 &0.1791
															&0.0395 &0.0625 &0.0826 &0.1724 \\ \hline
\multicolumn{1}{l|}{PRME-G} &0.0422 &0.065 &0.0813 & 0.1868
															&0.0466 &0.0723 &0.0876 &0.1715 \\ \hline
\multicolumn{1}{l|}{GE} 	&0.0294 &0.0329 &0.0714 &0.1691
															&0.0062 &0.0321 &0.0607 &0.1102\\ \hline
\multicolumn{1}{l|}{RNN} 	&0.0475 &0.0901 &0.1138 &0.1901
															&0.1321 &0.1867 &0.2043 &0.2186 \\ \hline
\multicolumn{1}{l|}{LSTM}   &0.0486 &0.0937 &0.1276 &0.1975
															&0.1261 &0.1881 &0.2019	&0.2123\\ \hline
\multicolumn{1}{l|}{GRU} 	&0.0483 &0.0915 &0.1216 &0.1934
															&0.1237 &0.1921 &0.1992 &0.2101 \\ \hline
\multicolumn{1}{l|}{ST-RNN} &0.0505 &0.0922 &0.1232 &0.2075
															&0.1379 &0.1957 &0.2091 &0.2239\\ \hline
\multicolumn{1}{l|}{ST-LSTM} &0.0716 &0.1232 &0.1508 &0.2208
															&0.1978 &0.2436 &0.2651 &0.3194\\ \hline
\multicolumn{1}{l|}{ST-CLSTM} &\textbf{0.0801} &\textbf{0.1308} &\textbf{0.1612}   &\textbf{0.2556} &\textbf{0.2037} &\textbf{0.2542} &\textbf{0.2861}  &\textbf{0.3433}\\ \hline  \hline

\multirow{2}{*}{} & \multicolumn{4}{c|}{\textbf{Gowalla}} & \multicolumn{4}{c}{\textbf{Brightkite}} \\ \cline{2-9}
& \multicolumn{1}{c|}{\textbf{Acc@1}} & \multicolumn{1}{c|}{\textbf{Acc@5}} & \multicolumn{1}{c|}{\textbf{Acc@10}} & \multicolumn{1}{c|}{\textbf{MAP}} & \multicolumn{1}{c|}{\textbf{Acc@1}} & \multicolumn{1}{c|}{\textbf{Acc@5}} & \multicolumn{1}{c|}{\textbf{Acc@10}} & \multicolumn{1}{c}{\textbf{MAP}} \\ \hline
\multicolumn{1}{l|}{FPMC-LR} &0.0293 &0.0524 &0.0849 &0.1745
															&0.1634 &0.2475 &0.3164 &0.33\\ \hline
\multicolumn{1}{l|}{PRME-G} & 0.0334 &0.0652 &0.0869 &0.1916
															&0.1976 &0.2993 &0.3495 &0.3115\\ \hline
\multicolumn{1}{l|}{GE} 	&0.0174 &0.06 &0.0947 	&0.1973
															&0.0521 &0.1376 &0.2118 &0.2602\\ \hline
\multicolumn{1}{l|}{RNN} 	&0.0473 &0.0892 &0.1207 &0.1998
															&0.3401 &0.4087 &0.432 &0.413 \\ \hline
\multicolumn{1}{l|}{LSTM} 	&0.0503 &0.0967 &0.1241 &0.2004
															&0.3575 &0.4146 &0.4489 &0.4303\\ \hline
\multicolumn{1}{l|}{GRU} 	&0.0498 &0.0931 &0.1289 &0.2045
															&0.331 &0.4007 &0.4377 &0.4042 \\ \hline
\multicolumn{1}{l|}{ST-RNN} &0.0519 &0.09532 &0.1304 &0.2187
															&0.3672 &0.4231 &0.4477 &0.4369\\ \hline
\multicolumn{1}{l|}{ST-LSTM}&0.0713 &0.1355 &0.1669 &0.2338
															&0.4389 &0.4807 &0.5035  &0.5266\\ \hline
\multicolumn{1}{l|}{ST-CLSTM} &\textbf{0.0778} &\textbf{0.1492} &\textbf{0.1818}  &\textbf{0.2557} &\textbf{0.4443} &\textbf{0.4953} &\textbf{0.5231} &\textbf{0.5626}\\ \hline
\end{supertabular}
\label{table:comparisons}
\vspace{-0.3cm}
\end{table*}

\textbf{Method Comparison.} The performance of our proposed model ST-LSTM and the seven baselines on four datasets evaluated by Acc@K and MAP is shown in Table 2. The cell size and the hidden state size are set as 128 in our experiments. The number of Epochs is set as 100 and the batch size is set as 10 for our proposed model. Other baseline parameters follow the best settings in their papers. From the experimental results, we can see following observations:
RNN performs better than Markov chain method FPMC-LR and embedding method PRME-G, due to its capability in modeling sequential data and user interests using RNN cell. Both LSTM and GRU slightly improve the performance compare with RNN because of their advantages in modeling long-term interests.
The result of GE is not good for missing social and textual information in our datasets.
The performance of the state-of-the-art method ST-RNN is close to the standard RNN method, which may be caused by the difficulty of manually setting the windows of time and distance intervals.
Another reason may be that the setting of the window does not well model the relation of recently visited POIs and next POI.
Our model ST-LSTM outperforms all baselines on the four datasets. The significant improvement of ST-LSTM indicates that it can well model temporal and spatial contexts. This is because we add time and distance gates to integrate time and distance intervals into the model. Moreover, ST-CLSTM not only reduces the number of parameters, but also marginally improve the performance compared with ST-LSTM.

\textbf{Effectiveness of Time and Distance Gates.} There are two time gates and two distance gates in our ST-CLSTM model.
We first investigate the effectiveness of time and distance gates on modeling time and distance intervals.
Specifically, we set $D1_t = 1$ and $D2_t  = 1$, in Eq.~(9) and Eq.~(10), respectively. That is, we close two distance gates and only consider the time intervals. Similarly, we set $T1_t = 1$ and $T2_t  = 1$, in Eq.~(7) and Eq.~(8), respectively. That is, we close two time gates and only consider distance information. From Figure~4, we can observe that the time gates and distance gates have almost equal importance on the two datasets (i.e., Gowalla and CA). Moreover, they both are critical for improving the recommendation performances.

We also investigate the effectiveness of time and distance gates on modeling short-term and long-term interests.
We set $T2_t = 1$ and $D2_t  = 1$, in Eq.~(8) and Eq.~(10), which means we close time and distance gates on long-term interests and only activate time and distance gates on short-term interest. Similarly, we set $T1_t = 1$ and $D1_t  = 1$, in Eq.~(7) and Eq.~(9), which means we close time and distance gates for short-term interest. As shown in Figure~4,  we can observe that they all perform worse than original ST-CLSTM, which means that time and distance intervals are not only critical to short-term interests but also important to long-term interests. Distance intervals may help model user general spatial preference and time intervals may help to model user long-term periodical behavior.
\begin{figure}[!htp]
\vspace{-0.3cm}
\setlength{\abovecaptionskip}{0.4cm}
\setlength{\belowcaptionskip}{-0.1cm}
\centering
\begin{minipage}{0.45\linewidth}
    \subfigure[Gowalla - Acc@K]{
      \label{experiment1:a} 
      \includegraphics[width=4.3cm,height=3cm]{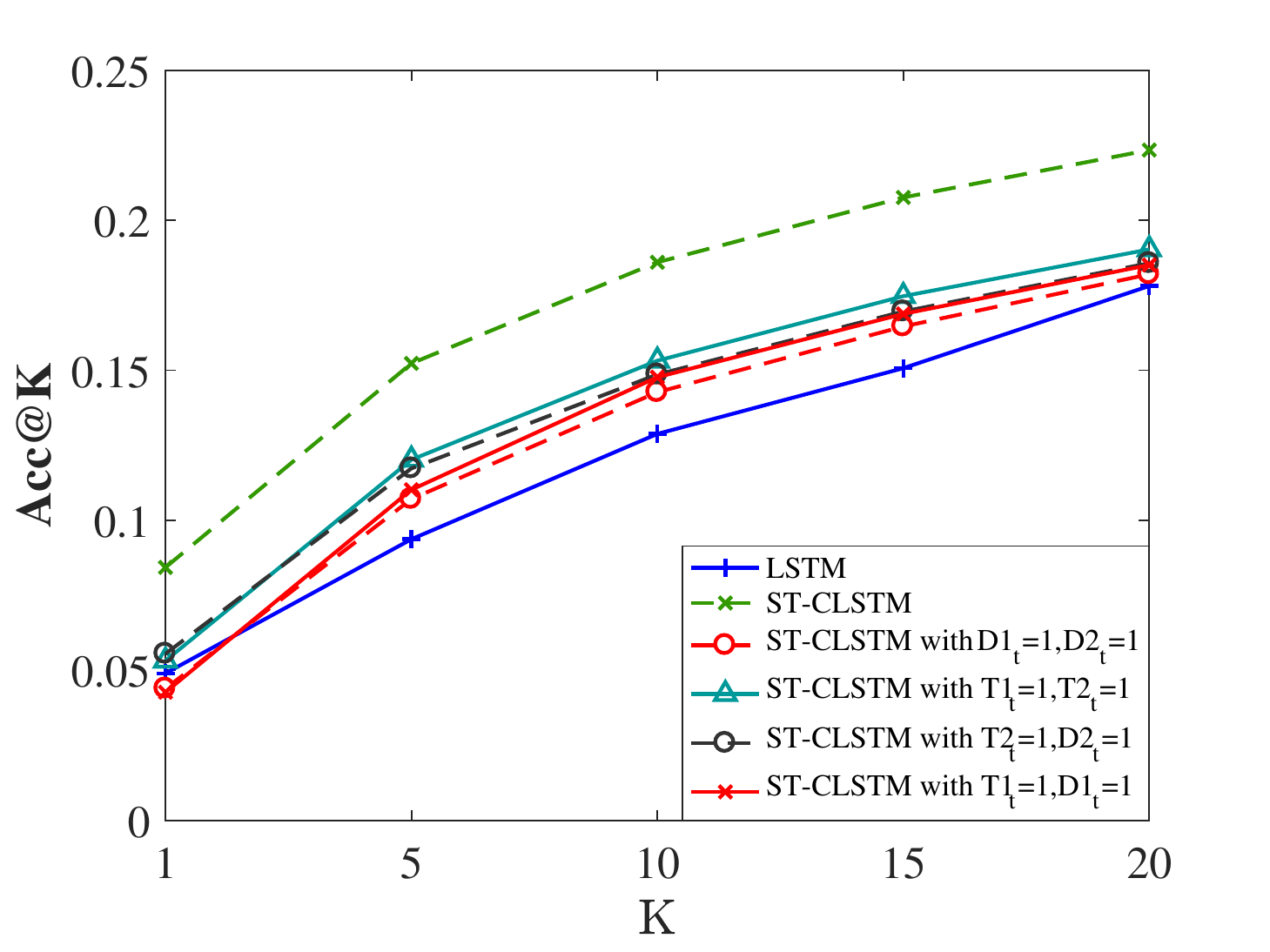}}
\end{minipage}
\begin{minipage}{0.45\linewidth}
	\subfigure[CA - Acc@K]{
      \label{experiment2:a} 
      \includegraphics[width=4.3cm,height=3cm]{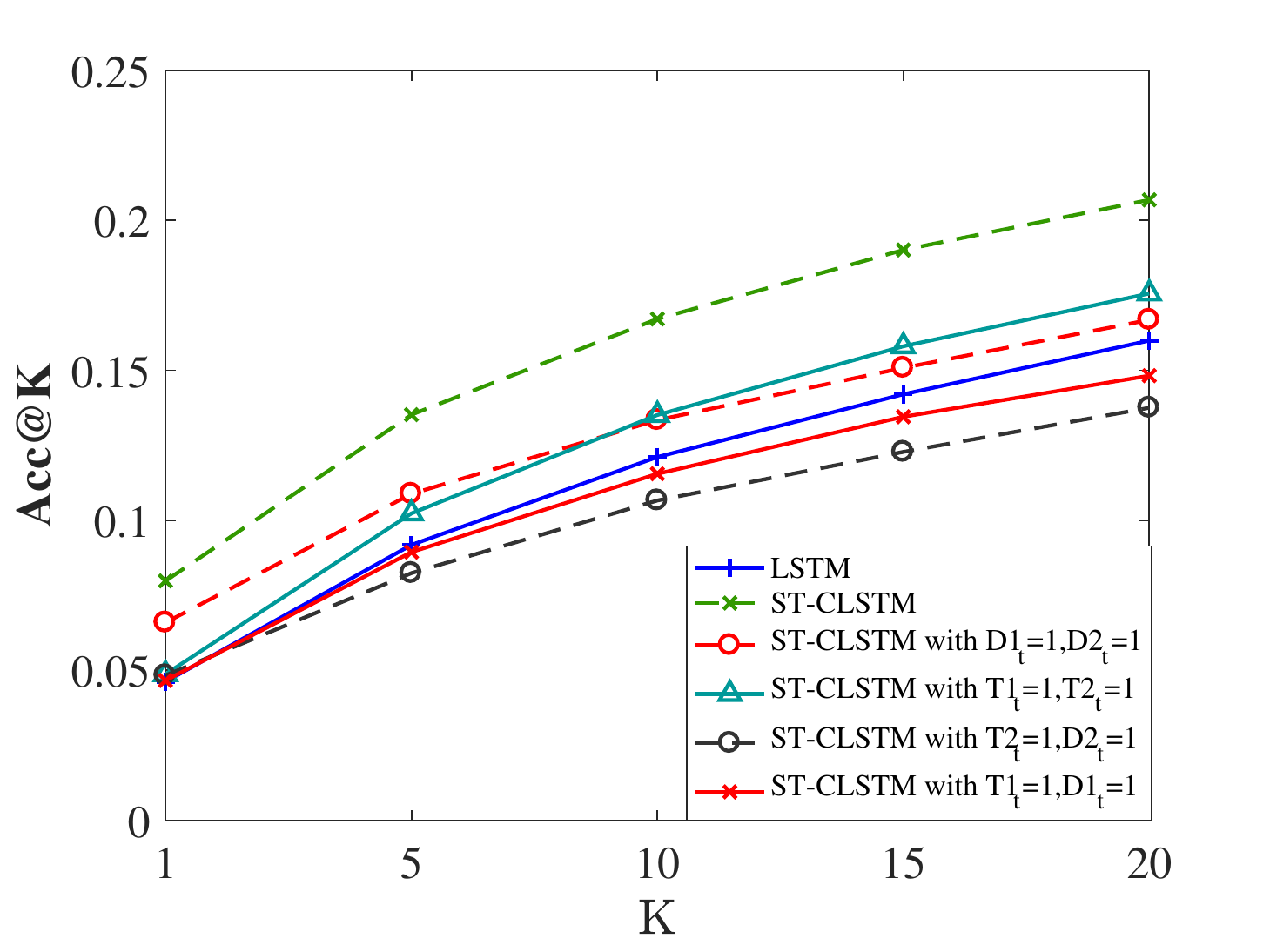}}
   \label{experiment2}
\end{minipage}
\vspace{-0.1cm}
\caption{The performance with different time and distance gates in ST-CLSTM}
\label{fig:gates}
\vspace{0.1cm}
\end{figure}

\textbf{Performance of Cold Start.}
We also evaluate the performance of ST-LSTM by comparing with other competitors for cold-start users. If a user just visits a few POIs, we think the user is cold. Specifically, we take users with less than 5 check-ins as a cold user in our experiments. We conduct the experiments on two datasets (i.e., Gowalla and BrightKite) and use Acc@K as the measure metric. As shown in Figure 5, we can observe that ST-CLSTM performs the best among all methods under cold start scenario. The reason is that ST-CLSTM models long-term interests as well as short-term interests with considering time and distance intervals.

\begin{figure}[!htp]
\vspace{-0.3cm}
\setlength{\abovecaptionskip}{0.4cm}
\setlength{\belowcaptionskip}{-0.1cm}
\centering
\begin{minipage}{0.45\linewidth}
    \subfigure[Gowalla - Acc@K]{
      \label{experiment1:a} 
      \includegraphics[width=\textwidth]{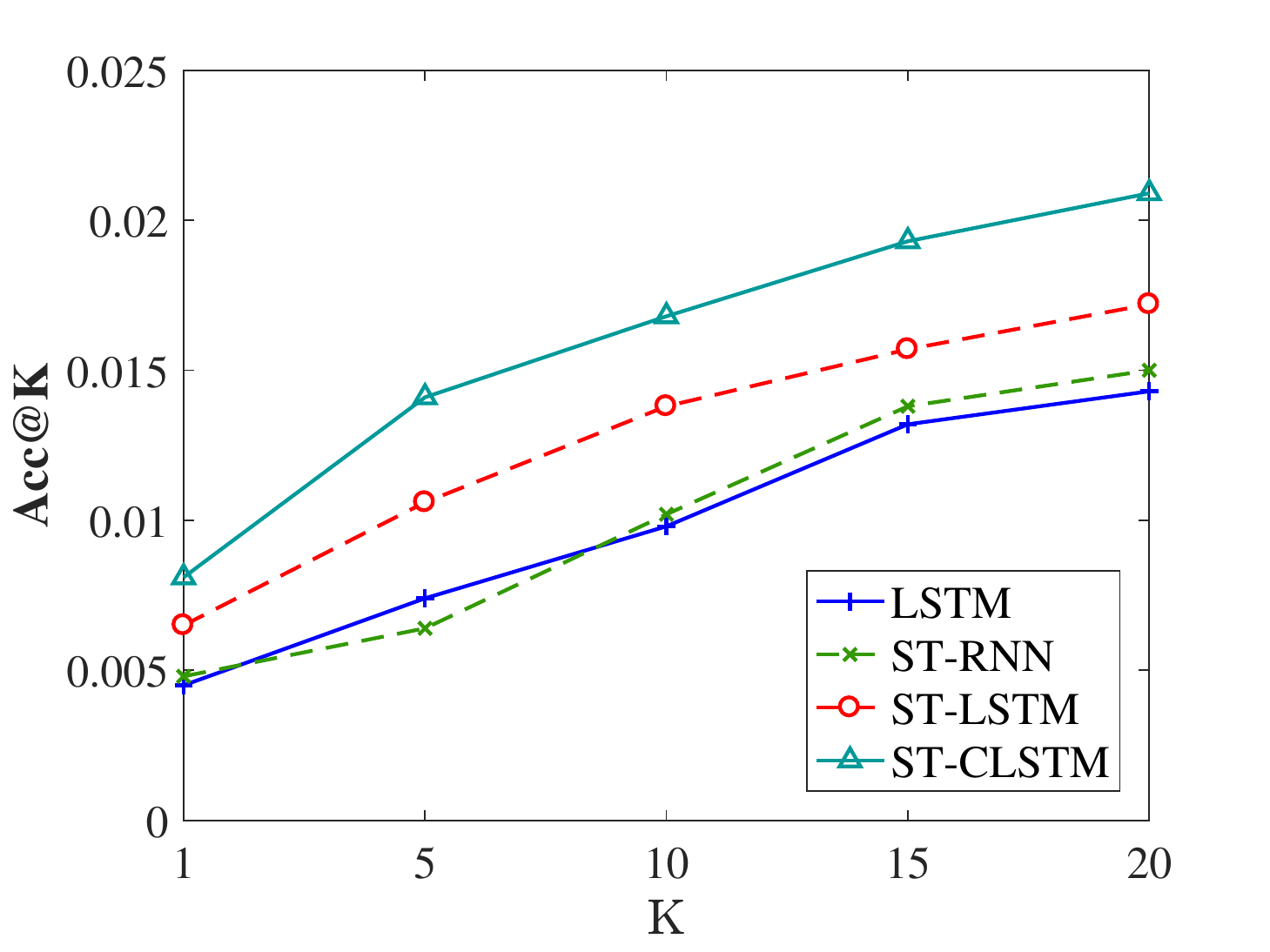}}
\end{minipage}
\begin{minipage}{0.45\linewidth}
	\subfigure[BrightKite - Acc@K]{
      \label{experiment2:a} 
      \includegraphics[width=\textwidth]{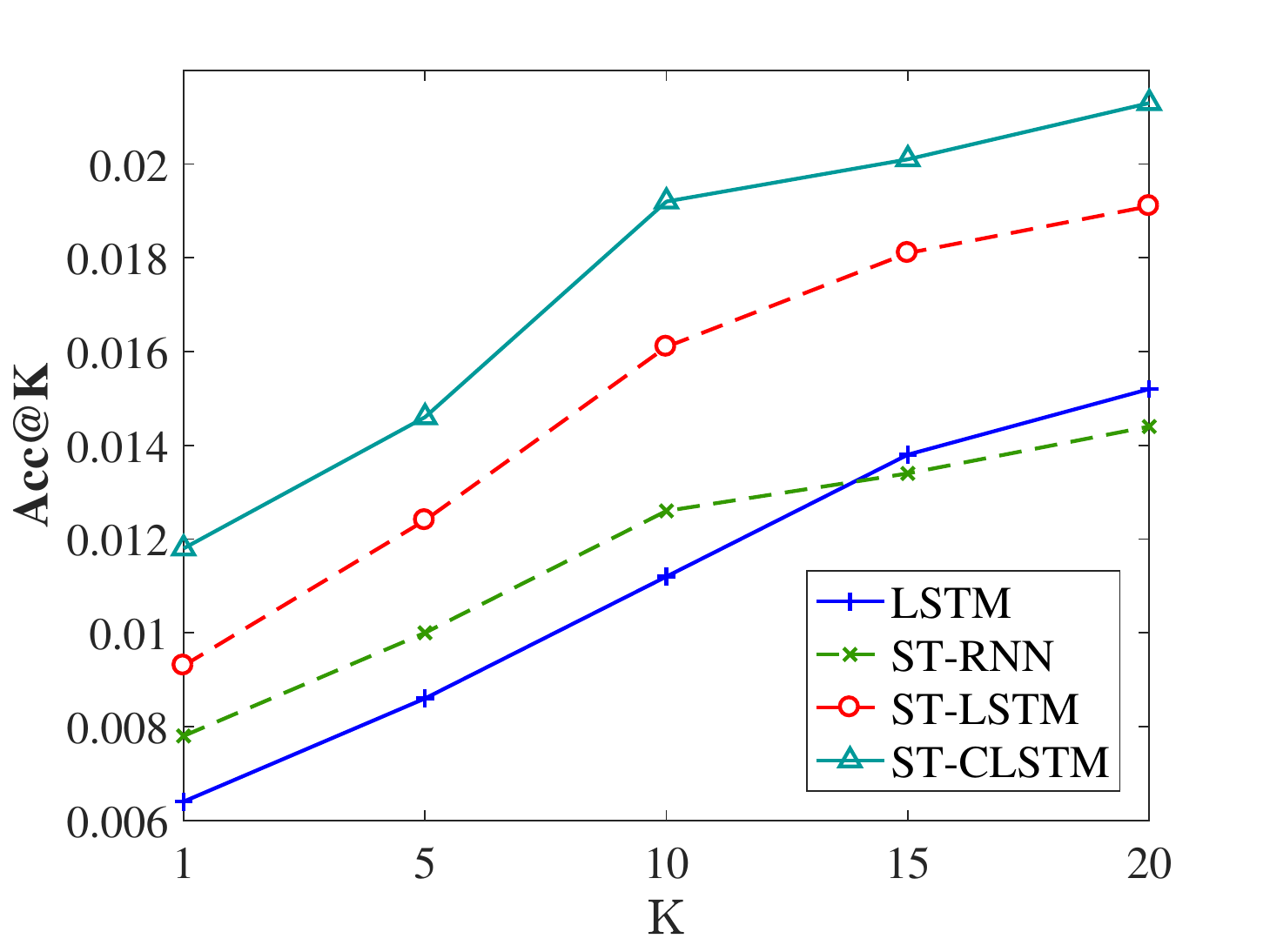}}
   \label{experiment2}
\end{minipage}
\vspace{-0.1cm}
\caption{The performance of cold start on two datasets}
\label{fig:cold start}
\vspace{-0.1cm}
\end{figure}

\textbf{Impact of Parameters.}
In the standard RNN, different cell sizes and batch sizes may lead to different performances. We investigate the impact of these two parameters for ST-LSTM and ST-CLSTM. We vary cell sizes and batch sizes to observe the performance and the training time of our proposed two models. We only show the impact of the two parameters on Gowalla dataset due to space constraint. As shown in Figure 6, increasing the cell size can improve our model in terms of the Acc@10 metric, and a proper batch size can help achieve the best performance. The cell size determines the model complexity, and the cell with a larger size may fit the data better. Moreover, a small batch size may lead to local optimum, and a big one may lead to insufficient updating of parameters in our two models.
\begin{figure}[!htp]
\vspace{-0.3cm}
\setlength{\abovecaptionskip}{0.4cm}
\setlength{\belowcaptionskip}{-0.1cm}
\centering
\begin{minipage}{0.45\linewidth}
    \subfigure[Different Cell Size]{
      \label{experiment1:a} 
      \includegraphics[width=4cm,height=3cm]{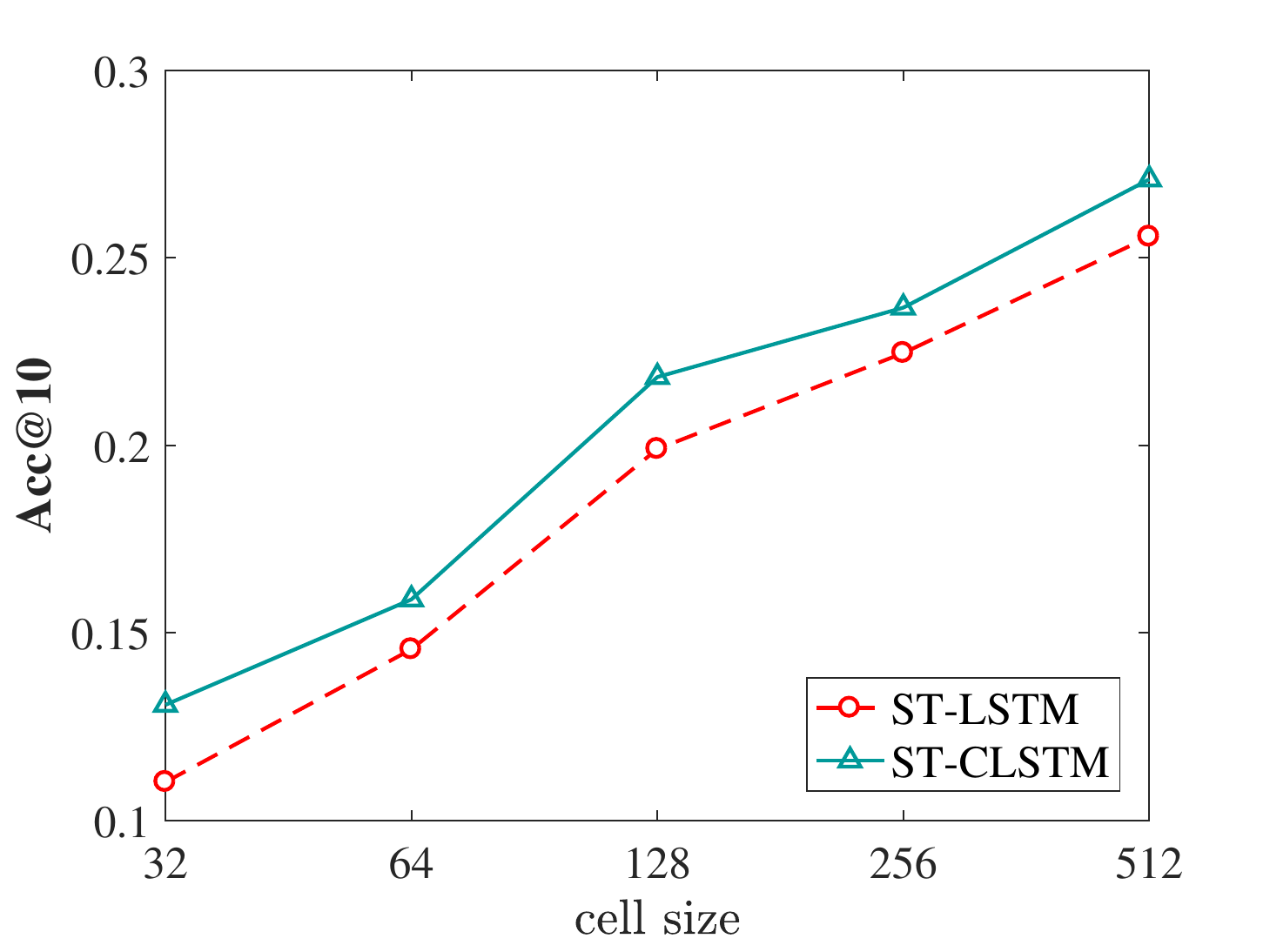}}
\end{minipage}
\begin{minipage}{0.45\linewidth}
	\subfigure[Different Batch Size]{
      \label{experiment2:a} 
      \includegraphics[width=4cm,height=3cm]{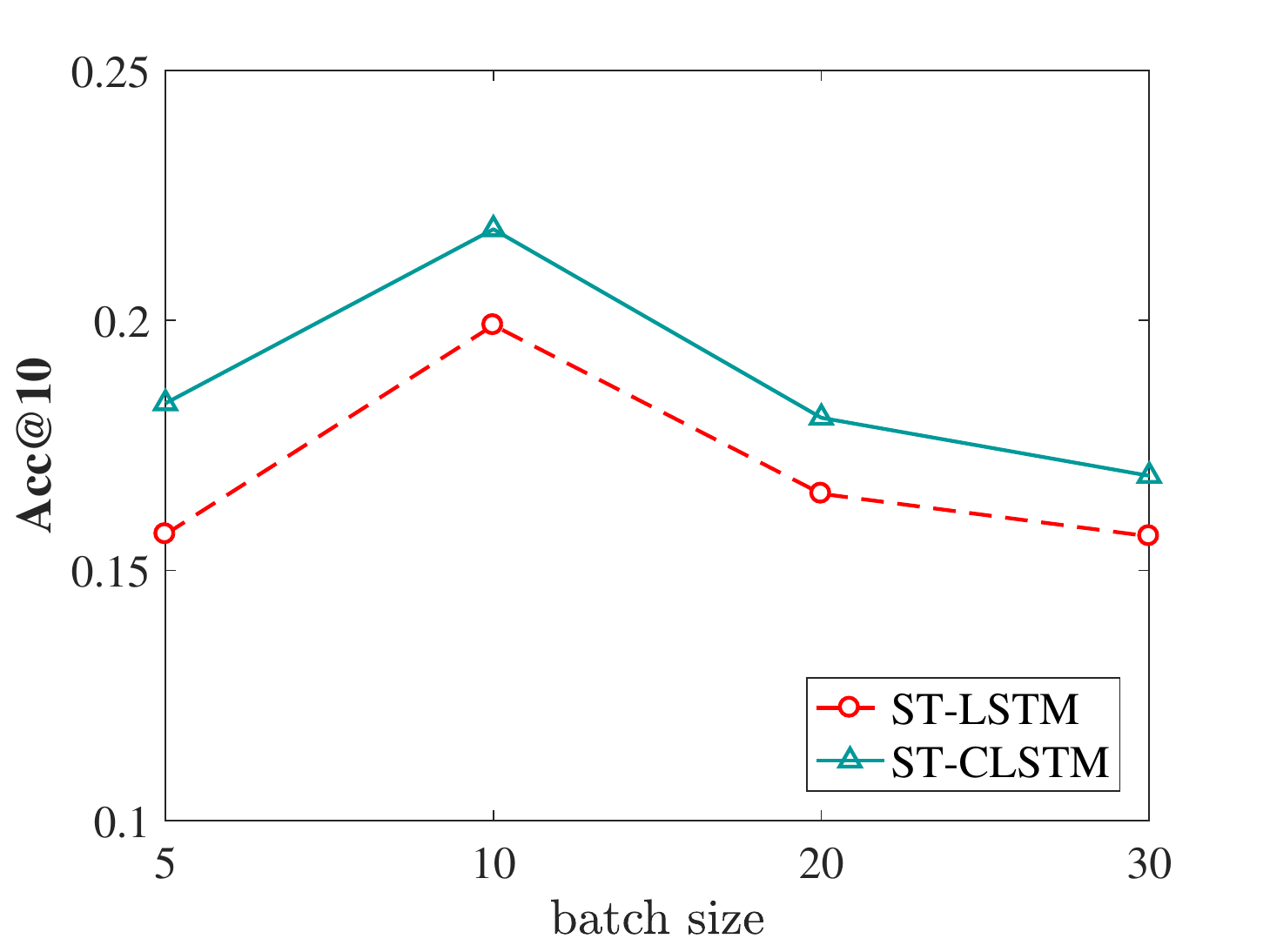}}
   \label{experiment2}
\end{minipage}
\vspace{-0.1cm}
\caption{The performance with different cell sizes and batch sizes on Gowalla}
\label{fig:cell size}
\vspace{-0cm}
\end{figure}

\section{Conclusions}
In this paper, a spatio-temporal recurrent neural network, named ST-LSTM, was proposed for next POI recommendation.
Time and distance intervals between neighbor check-ins were modeled using time and distance gates in ST-LSTM.
Specifically, we added a new cell state, and so there are two cell states to memorize users' short-term and long-term interests respectively. We designed time and distance gates to control user's short-term interest update and another pair of gates to control long-term interest update, so as to improve next POI recommendation performance.
We further coupled time and distance gates to improve ST-LSTM efficiency.
Experimental results on four large-scale real-world datasets demonstrated the effectiveness of our model, which performed better than the state-of-the-art methods.
In future work, we would incorporate more context information such as social network and textual description content into the model to further improve the next POI recommendation accuracy.

\bibliographystyle{named}
\bibliography{ijcai18}

\end{document}